\documentclass[preprintnumbers,amsmath,amssymb]{revtex4}
\usepackage{epsfig}
\usepackage{color}
\usepackage[colorlinks,urlcolor=blue,linkcolor=blue,citecolor=red]{hyperref}
\begin{document}
\setlength{\voffset}{1.0cm}
\title{SU($N$) affine Toda solitons and breathers from transparent Dirac potentials}
\author{Michael Thies\footnote{michael.thies@gravity.fau.de}}
\affiliation{Institut f\"ur  Theoretische Physik, Universit\"at Erlangen-N\"urnberg, D-91058, Erlangen, Germany}
\date{\today}

\begin{abstract}
Transparent scalar and pseudoscalar potentials in the one-dimensional Dirac equation play an important role as self-consistent
mean fields in 1+1 dimensional four-fermion theories (Gross-Neveu, Nambu--Jona Lasinio models) and quasi-one dimensional
superconductors (Bogoliubov-De Gennes equation). Here, we show that they also serve as seed to generate a large class of
classical multi-soliton and multi-breather solutions of su($N$) affine Toda field theories, including the Lax representation and the corresponding
vector. This generalizes previous findings about the relationship between real kinks in the Gross-Neveu model and classical solitons of
the sinh-Gordon equation to complex twisted kinks.
\end{abstract}
\maketitle
\section{Introduction}
\label{sect1}

Transparent potentials of one-dimensional, quantum mechanical wave equations have a vanishing reflection coefficient at all energies.
They are closely related to explicitly solvable soliton problems and integrability, as most clearly seen in the inverse scattering problem \cite{1a,1b,1c}.
Specifically, the case
which interests us here is the Dirac equation in one space dimension with Lorentz scalar ($S$) and pseudo-scalar ($P$) potentials,   
\begin{equation}
( i \partial \!\!\!/ - S - i \gamma_5 P)\psi = 0.
\label{1.1}
\end{equation}
By choosing a chiral representation of the Dirac matrices (diagonal $\gamma_5$)
\begin{equation}
\gamma^0 = \sigma_1, \quad \gamma^1 = i \sigma_2, \quad \gamma_5 = \gamma^0 \gamma^1 = - \sigma_3,
\label{1.2}
\end{equation}
together with light cone coordinates
\begin{equation}
z=x-t, \quad \bar{z}=x+t, \quad \partial_0 = \bar{\partial}-\partial, \quad \partial_1 = \bar{\partial}+\partial, \quad \square = - 4 \partial \bar{\partial},
\label{1.3}
\end{equation}
the Dirac equation (\ref{1.1}) simplifies to
\begin{equation}
2i \partial \psi_L = - \Delta^* \psi_R, \quad
2i \bar{\partial} \psi_R = \Delta \psi_L .
\label{1.4}
\end{equation}
We have introduced the complex potential
\begin{equation}
\Delta = S-iP
\label{1.5}
\end{equation}
and labeled the upper and lower spinor components according to their chirality as $\psi_L$ and $\psi_R$.

A large class of static and time-dependent potentials for Eq.~(\ref{1.4}) with vanishing reflection coefficient at all energies is known in closed, 
analytic form \cite{1,2}. The motivation to study such potentials actually came from 1+1 dimensional quantum field theories with four-fermion interaction (Gross-Neveu (GN) \cite{3} and
Nambu--Jona-Lasinio (NJL$_2$) \cite{4} models), as well as from the Bogoliubov-de Gennes (BdG) equation for quasi one-dimensional superconductors \cite{5,6}. Here, transparent
potentials emerge dynamically as mean fields in a semi-classical approximation to interacting fermion theories. 
They exhibit a variety of static solitons and non-static breathers, as well as bound states and time dependent scattering processes thereof \cite{7,8,9}.
The elementary building block of all of these solutions is the twisted kink carrying a topological charge, originally introduced by Shei in the large $N$ limit of the NJL$_2$
model \cite{10}.``Twist" denotes the complex phase factor between the vacuum values of $\Delta$ at $x \to \pm \infty$, here parametrized by a twist angle
$\varphi \in [0,\pi]$,
\begin{equation}
\lim_{x \to \infty} \Delta = e^{-2i\varphi} \lim_{x \to -\infty} \Delta.
\label{1.5a}
\end{equation}
These solitons are populated by fermions, but otherwise behave exactly as expected for integrable bosonic systems: Forward scattering only,
purely elastic scattering and factorization of the transmission amplitude into two-soliton transmission amplitudes. The fact that the bosons are composite rather than 
elementary does not affect any of these properties.

A question which has only been answered partially so far is the following: Do these transparent potentials satisfy one or several non-linear partial
differential equations (PDE's), which would allow one to relate the fermionic models to classical bosonic field theories? The only case where this has
been understood is dynamics of real kinks in the GN model ($P=0$, twist angle $\pi/2$ only, kink interpolating between the vacua $S=1$ to $S=-1$
in natural units) \cite{11}. This problem is much easier, since the fermion bound states do not react back onto the potential, and real kinks form neither bound states nor breathers.
It has been shown that kink dynamics (restricting twist to the value $-1$) in the large $N$ limit  can be mapped onto
the classical $N=2$ GN model, for which Neveu and Papanicolaou have proven long ago \cite{12} that $\ln S^2$ satisfies the sinh-Gordon equation,
\begin{equation}
\square \theta + 4 \sinh \theta = 0, \quad S^2=e^{\theta}.
\label{1.6}
\end{equation}
Notice that although $\theta$ cannot be interpreted as soliton solution of the sinh-Gordon equation (which do not exist, unlike in the sine-Gordon case),
$S=e^{\theta/2}$ behaves in all respects like one or several solitons.
The Lax pair used to prove integrability is given by
\begin{equation}
\partial \psi = {\cal A} \psi, \quad \bar{\partial} \psi = {\cal B} \psi,
\label{1.7}
\end{equation}
with 
\begin{equation}
{\cal A} = \frac{1}{2i} \left( \begin{array}{cc} 0 & - e^{\theta/2} \\ - \zeta^{-2}e^{-\theta/2}  &  i \partial \theta \end{array} \right),
\quad
{\cal B} = \frac{1}{2i} \left( \begin{array}{cc} i \bar{\partial} \theta & \zeta^2 e^{-\theta/2} \\ e^{\theta/2} & 0  \end{array} \right),
\label{1.8}
\end{equation}
in a certain gauge ($\zeta$ is a spectral parameter, cf. Eq. (\ref{2.2}) below). The integrability condition for the linear system (\ref{1.7}) (or zero curvature condition)
\begin{equation}
\bar{\partial} {\cal A} - \partial {\cal B} + \left[ {\cal A},{\cal B} \right] = 0
\label{1.9}
\end{equation}
then reproduces the sinh-Gordon equation (\ref{1.6}).
In the gauge used here, the spinors $\psi$ in Eq.~(\ref{1.7}) are the Dirac spinors with components $\psi_L, \psi_R$. Indeed, Eqs.~(\ref{1.7},\ref{1.8}) encompass the Dirac equation (\ref{1.4}) 
for $\partial \psi_L$ and $\bar{\partial} \psi_R$, but also two further linear equations for $\partial \psi_R$ and $\bar{\partial} \psi_L$. It was therefore referred to 
as ``extended Dirac equation" in Ref.~\cite{10}.

Actually, in this case, since multi-soliton solutions of the sinh-Gordon equation were already well known \cite{13}, they could be used to solve the 
many-kink scattering problem in full generality in the large $N$ GN model \cite{14}. As a by-product, this made it possible to relate GN kink
dynamics to folded strings in AdS$_2$ \cite{15}, an observation which has subsequently found a simple geometrical interpretation in Ref.~\cite{16}. 

If we allow for twists
different from ($-1$), we are now just in the opposite situation. We have control over soliton dynamics in fermionic models, but cannot identify corresponding 
bosonic nonlinear PDE's. Such an identification would hopefully provide us with a deeper understanding of the soliton solution (found by ansatz so far) and
let us benefit from the sophisticated mathematical machinery which has been developed for integrable bosonic models over the years.

In the present work, we present the solution to this problem. Unlike Neveu and Papanicolaou, we have not yet succeeded in deriving the PDE's directly 
from a fermionic field theory. We rather start from the solution for transparent potentials $\Delta$ and try to find PDE's which they satisfy. This has turned out to be
less straightforward than it may sound. The reason is the fact that the answer involves coupled PDE's for a number of scalar fields (even infinitely many in the
most general case), whereas 
the fermionic theory provides us only with a single complex field $\Delta$. To correctly identify the other fields has been the main obstacle 
preventing us from finding the answer in the past. 

This paper is organized as follows. In Sect.~\ref{sect2}, we summarize the known results for transparent Dirac potentials which are needed later on. 
In Sect.~\ref{sect3} we show how to generate out of these results soliton solutions of classical affine Toda field theory related to the Lie algebra su($\infty$). 
Sect.~\ref{sect4} is dedicated to multi-soliton solutions of su($N$) Toda models for finite $N$. Since the sine-Gordon model is recovered for $N=2$,
this seems to be the natural generalization of the previous work to twists different from ($-1$). In Sect.~\ref{sect5} we comment on how to include
breathers into this framework. In the concluding Sect.~\ref{sect6}, we compare our results with known classical Toda soliton solutions and give a brief 
outlook.

\section{Reminder of transparent Dirac potentials}
\label{sect2}
Here we summarize the results for all known transparent Dirac potentials $\Delta$, see Eq.~(\ref{1.5}), in the notation of Ref.~\cite{2}. 
Our starting point is the Dirac equation (\ref{1.1}-\ref{1.4}).
We shall also need the continuum spinors
\begin{equation}
\psi_{\zeta} = \left(\begin{array}{c} \psi_L \\ \psi_R \end{array} \right) =   \frac{1}{\sqrt{1+\zeta^2}} \left( \begin{array}{c} \zeta \chi_1 \\ - \chi_2 \end{array} \right) e^{i(\zeta \bar{z}-z/\zeta)/2} 
\label{2.1}
\end{equation}
belonging to the potential $\Delta$, labeled by a spectral parameter (fermion mass $m=1$)
\begin{equation}
k = \frac{1}{2} \left( \zeta-\frac{1}{\zeta} \right), \quad E=\pm \sqrt{k^2+1} = - \frac{1}{2} \left( \zeta+ \frac{1}{\zeta} \right).
\label{2.2}
\end{equation}
The functions $\chi_1, \chi_2$ are defined in such a way that they reduce to $\chi_1=\chi_2=1$ for the free theory ($\Delta=1$).
If one expresses the spectral parameter $\zeta$ in terms of $k,E$ and the light cone coordinates $z, \bar{z}$ in terms of ordinary coordinates $t,x$,
the exponential factor in (\ref{2.1}) assumes the more familiar form $e^{i(kx-Et)}$.

It turns out that the most compact way of writing down the multi-soliton, multi-breather solution is in terms of the following ratios of determinants,
\begin{eqnarray}
\Delta & = & \frac{{\rm det}(\omega+A)}{{\rm det}(\omega + B)},
\nonumber \\
\chi_1 & = & \frac{{\rm det}(\omega+C)}{{\rm det}(\omega + B)},
\nonumber \\
\chi_2 & = & \frac{{\rm det}(\omega+D)}{{\rm det}(\omega + B)}.
\label{2.3}
\end{eqnarray}
We have introduced 5 square matrices $\omega, A, B, C, D$ of dimension $n \times n$ for a general solution with $n$ bound states,
built out of $n$ elementary twisted kinks. The most basic matrix is $B$ appearing in all three denominators and defined as follows,
\begin{eqnarray}
B_{k \ell} & = & i \frac{e_k e_{\ell}^*}{\zeta_{\ell} - \zeta_{k}^*}, 
\nonumber \\
e_k & = & e^{i (\zeta_k^*\bar{z}-z/\zeta_k^*)/2},
\nonumber \\
\zeta_k & = &  - e^{-i \varphi_k- \xi_k}.
\label{2.4}
\end{eqnarray} 
The complex spectral parameter $\zeta_k$ encodes the chiral twist angle $\varphi_k$ and the rapidity (velocity $v_k = \tanh  \xi_k$)
of the $k$-th twisted kink and corresponds to the position of a bound state pole in the complex $\zeta$-plane. Let us furthermore introduce the 
diagonal $n \times n$ matrix $Z$ as
\begin{equation}
Z_{k \ell}= \delta_{k \ell} \zeta_k \quad ({\rm no\ }k {\rm \ sum}).
\label{2.5}
\end{equation}
Then $A,C,D$ can be related to $B$ as follows,
\begin{eqnarray}
A & = &  (Z^{\dagger})^{-1} B Z,
\nonumber \\
C & = & (\zeta- Z^{\dagger}) B (\zeta- Z)^{-1},
\nonumber \\
D & = & (Z^{\dagger})^{-1} C Z.
\label{2.6}
\end{eqnarray}
Finally, $\omega$ is a hermitean, constant matrix determining positions and initial conditions of solitons and
details of breathers, in case $\omega$ is non-diagonal. The ratios of determinants can actually be computed
and expressed through the matrix $(\omega+B)^{-1}$. To this end, one uses the following identity for the
determinant of a matrix which is the sum of an invertible matrix and a one-term separable one \cite{17},
\begin{equation}
\frac{{\rm det}(M+a b^{\dagger})}{{\rm det}(M)} = 1 + b^{\dagger} \frac{1}{M} a.
\label{2.7}
\end{equation} 
Introducing the $n$-dimensional vector $e$ with components $e_k$, one finds
\begin{eqnarray}
\Delta & = & 1+i e^{\dagger} \frac{1}{\omega+B}\frac{1}{Z^{\dagger}} e,
\nonumber \\
\chi_1 & = & 1 + i e^{\dagger} \frac{1}{\zeta-Z} \frac{1}{\omega+B} e ,
\nonumber \\
\chi_2 & = &  1 +i \zeta e^{\dagger} \frac{1}{\zeta-Z} \frac{1}{\omega+B} \frac{1}{Z^{\dagger}} e.
\label{2.8}
\end{eqnarray}
This representation is useful for practical calculations and for proving the Dirac equation. We refer to
Ref.~\cite{2} for the actual proof. The determinental form is better suited for deriving the Lax
pair, see next section. 

A last comment concerns bound states. The above solution was actually found in Ref.~\cite{2} by starting
from the $n$ bound states of the $n$ soliton solution. The bound state spinors, in turn, were obtained by solving
algebraic systems of linear equations. The bound state spinors can also be retrieved from the  reflectionless
continuum spinors $\psi_{\zeta}$, namely as residues at the poles in the complex $\zeta$-plane,
\begin{equation}
\phi_{\ell} = \lim_{\zeta \to \zeta_{\ell}} (\zeta-\zeta_{\ell}) \psi_{\zeta}.
\label{2.15}
\end{equation}
They play an important role in applications of the present problem to BdG equation and GN models,
where they enter critically into the self-consistency condition. Since this is not our topic here and bound state spinors
can always be recovered from continuum spinors using (\ref{2.15}), we do not discuss them any further.

In Ref.~\cite{2}, the class of transparent Dirac potentials sketched above has been found by ansatz. To better understand
integrability issues, one would like to construct a Lax pair and derive a bosonic non-linear equation for the
transparent potential $\Delta$. The Dirac equation yields $\partial \psi_L$ and $\bar{\partial} \psi_R$ only.
If one tries to derive the missing equations for $\partial \psi_R$ and $\bar{\partial} \psi_L$, one discovers
that the equations do not close. It does not seem possible to express these derivatives through
$\psi_L, \psi_R, \Delta$ and derivatives of $\Delta$. This only worked for the case of real kinks (twist $-1$).
The only equation which could be derived for arbitrary twist is the so called ``master equation"
\begin{equation}
- 4 \partial \bar{\partial} \ln {\rm det\ } (\omega+B) +1 - |\Delta|^2 = 0.
\label{2.16}
\end{equation}
Eq.~(\ref{2.16}) can be derived in a few lines using the 
``log det = Tr log" identity, cf. Sect.~IIID of Ref.~\cite{2}.
It is evidently not sufficient to determine $\Delta$, but was spelled out in \cite{2}, hoping that it might give a clue
as to the full nonlinear theory.  In the next section, we will identify the
PDE's for general twist and put the master equation into the right context.

\section{Lax pair and relation to  su($\infty$) affine Toda field theory}
\label{sect3}

Let us insert the solution (\ref{2.1}-\ref{2.3}) into the Dirac equation (\ref{1.4}), using the identity
\begin{equation}
{\rm det}(\omega+(Z^{\dagger})^{-1} B Z) = {\rm det}( (Z^{\dagger})^{-1}Z) {\rm det} (Z^{\dagger} \omega Z^{-1} + B)
\label{3.1}
\end{equation}
(and similarly for $B \to C$) as well as the notation for the plane wave
\begin{equation}
e_{\zeta} = e^{i(\zeta \bar{z}-z/\zeta)/2}.
\label{3.2}
\end{equation}
This yields
\begin{eqnarray}
2 i \partial \left[ e_{\zeta} \frac{{\rm det}(\omega+C)}{{\rm det}(\omega+B)} \right] & = & \frac{e_{\zeta}}{\zeta}
\frac{{\rm det}((Z^{\dagger})^{-1} \omega Z + B) {\rm det}(Z^{\dagger}\omega Z^{-1} + C)}{{\rm det}(\omega+B)^2},
\nonumber \\
2i \bar{\partial} \left[ e_{\zeta} \frac{{\rm det}(Z^{\dagger} \omega Z^{-1} +C)}{{\rm det}(\omega+B)} \right] & = & - \zeta e_{\zeta}
\frac{ {\rm det}(Z^{\dagger}\omega Z^{-1}+B)   {\rm det}( \omega  + C)}{{\rm det}(\omega+B)^2}.
\label{3.3}
\end{eqnarray}
The various ways in which the matrix $\omega$ is dressed here suggests to introduce a whole tower of
non-hermitean generalizations of $\omega$, 
\begin{equation}
\omega_n = (Z^{\dagger})^n \omega Z^{-n} ,
\label{3.4}
\end{equation}
for integer $n$. The original $\omega$ coincides with $\omega_0$, and we read off the relationship 
\begin{equation}
\omega_{-n} = \omega_n^{\dagger}.
\label{3.5}
\end{equation}
Eq.~(\ref{3.3}) now becomes
\begin{eqnarray}
2i \partial \left[ e_{\zeta} \frac{{\rm det}(\omega_0+C)}{{\rm det}(\omega_0+B)} \right] & = & \frac{e_{\zeta}}{\zeta}
\frac{{\rm det}(\omega_{-1} + B) {\rm det}(\omega_1  + C)}{{\rm det}(\omega_0+B)^2},
\nonumber \\
2i \bar{\partial} \left[ e_{\zeta} \frac{{\rm det}( \omega_1 +C)}{{\rm det}(\omega_0+B)} \right] & = & - \zeta e_{\zeta}
\frac{ {\rm det}(\omega_1+B)   {\rm det}( \omega_0  + C)}{{\rm det}(\omega_0+B)^2}.
\label{3.6}
\end{eqnarray}
The crucial observation for our present goal is the following: 
Eqs.~(\ref{3.6}) are mathematical identities which hold for any matrix $\omega_0$, even a non-hermitean one. 
This can be inferred from the detailed proof of the Dirac equation in Ref.~\cite{2}. The fact that $\omega = \omega^{\dagger}$
for the original matrix $\omega$ only enters when we identify the potential $\Delta^*$ in the first Dirac equation (\ref{1.4}) as complex
conjugate of the potential $\Delta$ in the 2nd Dirac equation. If $\omega$ is non-hermitean, we are no longer able to
interpret (\ref{3.6}) as Dirac equation with scalar and pseudoscalar potentials, but it nevertheless remains valid.
Thus, in particular, Eq.~(\ref{3.6}) also holds if we start from $\omega_n$ (with arbitrary integer index $n$)
rather than $\omega$, i.e., substitute $\omega_0 \to \omega_n, \omega_1 \to \omega_{n+1}, \omega_{-1} \to \omega_{n-1}$,
\begin{eqnarray}
2i \partial \left[ e_{\zeta} \frac{{\rm det}(\omega_n+C)}{{\rm det}(\omega_n+B)} \right] & = & \frac{e_{\zeta}}{\zeta}
\frac{{\rm det}(\omega_{n-1} + B) {\rm det}(\omega_{n+1}  + C)}{{\rm det}(\omega_n+B)^2},
\nonumber \\
2i \bar{\partial} \left[ e_{\zeta} \frac{{\rm det}( \omega_{n+1} +C)}{{\rm det}(\omega_n+B)} \right] & = & - \zeta e_{\zeta}
\frac{ {\rm det}(\omega_{n+1}+B)   {\rm det}( \omega_n  + C)}{{\rm det}(\omega_n+B)^2}.
\label{3.7}
\end{eqnarray}
The structure of these equations becomes more transparent upon introducing two basic quantities as follows,
\begin{eqnarray}
\psi_n & = & e_{\zeta} \frac{{\rm det}(\omega_n+C)}{{\rm det}(\omega_n+B)},
\nonumber \\
e^{\Phi_n} & = & \frac{{\rm det}(\omega_n+B)}{{\rm det}(\omega_{n-1}+B)}.
\label{3.8}
\end{eqnarray}
This turns Eq.~(\ref{3.7}) into
\begin{eqnarray}
2i \zeta \partial \psi_n & = &  e^{-\Phi_n} \left( e^{\Phi_{n+1}}\psi_{n+1}\right),
\nonumber \\
2i \bar{\partial} \left(e^{\Phi_{n+1}} \psi_{n+1}\right) & = & - \zeta e^{\Phi_{n+1}} \psi_n.
\label{3.9}
\end{eqnarray}
The original Dirac equation can be recovered from this set of equations by setting $n=0$ and identifying
\begin{eqnarray}
\psi_0 = \frac{1}{\zeta} \psi_L, & \quad & \psi_1 = -  e^{-\Phi_1} \psi_R,
\nonumber \\
\Phi_1 = \ln \Delta, & \quad & \Phi_0 = -  \ln \Delta^*.
\label{3.10}
\end{eqnarray}
More importantly, both lines of Eq.~(\ref{3.9}) hold for arbitrary integer index $n$. In particular, we may now shift the 
index in the 2nd equation down by 1 and arrive at
\begin{eqnarray}
2i \zeta \partial \psi_n & = &  e^{\Phi_{n+1}-\Phi_n} \psi_{n+1},
\nonumber \\
2i \bar{\partial} \psi_n & = & - 2i (\bar{\partial}\Phi_n) \psi_n - \zeta  \psi_{n-1}.
\label{3.11}
\end{eqnarray} 
Eq.~(\ref{3.11}) has the form of a Lax pair for an infinite dimensional vector $\psi$ with
components $\psi_n, n=-\infty,...,\infty$,
\begin{equation}
\partial \psi = {\cal A} \psi, \quad \bar{\partial} \psi = {\cal B} \psi.
\label{3.12}
\end{equation}
with
\begin{eqnarray}
{\cal A}_{n,m} & = & \frac{1}{2i\zeta} \delta_{m,n+1} e^{\Phi_{n+1}-\Phi_n} \quad ({\rm no\ }n{\rm \ sum}),
\nonumber \\
{\cal B}_{n,m} & = & - \delta_{m,n} (\bar{\partial} \Phi_n )   - \frac{\zeta}{2i} \delta_{m,n-1}  \quad ({\rm no\ }n{\rm \ sum}).
\label{3.13}
\end{eqnarray}
The reduction of the infinite dimensional matrices ${\cal A},{\cal B}$ defined here to the two dimensional
matrices introduced in Eqs.~(\ref{1.7},\ref{1.8}) will be clarified below, see Eq.~(\ref{4.17}).
The price to pay for this Lax representation was the extension from a two-component Dirac spinor and a single complex potential 
to an infinite dimensional vector and infinitely many potentials, although the various components are related
to the original ones in a rather trivial fashion, see Eq.~(\ref{3.8}). The consistency condition for the linear system (\ref{3.12}) is
\begin{equation}
\bar{\partial} {\cal A} - \partial {\cal B} + [{\cal A},{\cal B}] = 0,
\label{3.13a}
\end{equation}
i.e., the zero-curvature condition characteristic for the Lax pair formulation of integrable systems. Applying $\bar{\partial}$ to the first line of Eq.~(\ref{3.11})
and $\partial$ to the 2nd line and using again Eq.~(\ref{3.11}) with shifted indices $n$, we finally arrive at
\begin{equation}
- 4 \partial \bar{\partial} \Phi_n =  \square \Phi_n = e^{\Phi_{n+1}-\Phi_n} - e^{\Phi_n- \Phi_{n-1}} \quad (n=-\infty,...,\infty).
\label{3.14}
\end{equation} 
These are the classical field equations of the affine Toda field theory corresponding to the limit $N\to \infty$ of
the Lie algebra A$_{N-1}^{(1)}$ or su($N$). (They differ from the original form of Mikhailov \cite{18} merely by scale factors
of 2 for the fields $\Phi_n$ and the light cone coordinates.) However, we have one additional condition, reducing the
number of independent bosonic fields by a factor of 2, since relation (\ref{3.5}) implies that
\begin{equation}
\Phi_{-n} = - \Phi_{n+1}^*.
\label{3.15}
\end{equation}
Note that owing to this symmetry, the classical Lagrangian from which (\ref{3.14}) can be derived,
\begin{equation}
{\cal L} = \frac{1}{2} \sum_{n=-\infty}^{\infty} \partial_{\mu} \Phi_n \partial^{\mu} \Phi_n - \sum_{n=-\infty}^{\infty} e^{\Phi_{n+1}- \Phi_n}
\label{3.16}
\end{equation}
is real, even though the fields $\Phi_n$ are complex:
\begin{equation}
{\cal L}  =  \frac{1}{2} \sum_{n=1}^{\infty} \left( \partial_{\mu} \Phi_n \partial^{\mu} \Phi_n+  \partial_{\mu} \Phi_n^* \partial^{\mu} \Phi_n^* \right)
- e^{\Phi_1+ \Phi_1^*} - \sum_{n=1}^{\infty} \left( e^{\Phi_{n+1}-\Phi_n} + e^{\Phi_{n+1}^*- \Phi_n^*} \right).
\label{3.17}
\end{equation}
Likewise, the Hamiltonian density is also real, but not bounded from below for complex fields $\Phi_n$.
The subset of independent Euler-Lagrange equations (\ref{3.14}) is
\begin{eqnarray}
- 4 \partial \bar{\partial} \Phi_1 & = &   e^{\Phi_2-\Phi_1} - e^{\Phi_1+ \Phi_1^*} ,
\nonumber \\
- 4 \partial \bar{\partial} \Phi_n & = &  e^{\Phi_{n+1}-\Phi_n} - e^{\Phi_n- \Phi_{n-1}} \quad (n=2,...,\infty).
\label{3.18}
\end{eqnarray} 
No obvious reduction of the Lax pair (\ref{3.11}) could be found, since the spinor components $\psi_n$ for positive and negative
indices $n$ are not related in any simple way.

Let us now come back to the master equation (\ref{2.16}). Using the definition of $\omega_n$ in Eq.~(\ref{3.4}), it can be written as
\begin{equation}
- 4 \partial \bar{\partial} \ln {\rm det\ }(\omega_0+B) + 1 - \frac{{\rm det\ }(\omega_{-1}+B) {\rm det\ }(\omega_1+B)}{{\rm det\ }(\omega_0+B)^2} = 0.
\label{3.19}
\end{equation}
If one goes through the derivation in Ref.~\cite{2}, one again finds that this equation holds for non-hermitean matrices $\omega_0$ as well.
This enables us once more to shift all indices on the $\omega$'s by an arbitrary integer, hence
\begin{equation}
- 4 \partial \bar{\partial} \ln {\rm det\ }(\omega_n+B) + 1 - \frac{{\rm det\ }(\omega_{n-1}+B) {\rm det\ }(\omega_{n+1}+B)}{{\rm det\ }(\omega_n+B)^2}= 0.
\label{3.20}
\end{equation}
Defining
\begin{equation}
e^{\phi_n} = {\rm det\ } (\omega_n+B),
\label{3.21}
\end{equation}
this is converted into
\begin{equation}
- 4 \partial \bar{\partial} \phi_n + 1 - e^{\phi_{n+1}-2\phi_n + \phi_{n-1}} = 0 ,\quad (n=-\infty,...,\infty).
\label{3.22}
\end{equation}
This equation differs from another standard formulation of affine su($\infty$) Toda field theory, see e.g. \cite{19}, only by the ``+1" term,
which can be traced back directly to a corresponding term in the master equation. Nevertheless, we can recover Eq.~(\ref{3.14}) by taking the 
difference between Eq.~(\ref{3.22}) for $n$ and $n-1$, defining 
\begin{equation}
\Phi_n = \phi_n- \phi_{n-1}.
\label{3.23}
\end{equation} 
The auxiliary condition (\ref{3.15}) now reads
\begin{equation}
\phi_{-n} = \phi_n^*.
\label{3.24}
\end{equation}
This derivation of the Toda field equations in Mikhailov's form \cite{18} is even more straightforward than the previous one and confirms the role of the master equation
as key to the Lax pair. The advantage of the derivation starting from the Dirac equation is the fact that we
get the Lax representation (\ref{3.13}), including the vector $\psi$, at the same time as the field equations. 

\section{Soliton solutions of su($N$) affine Toda field theories for arbitrary $N$}
\label{sect4}

So far, the twist angles $\varphi_k$ could take on any values between 0 and $\pi$, spanning the whole range of chiral twists of elementary kinks. 
Suppose now that we choose a positive integer $N \ge 2$ and discretize the twist angles in steps of $\pi/N$,
\begin{equation}
\varphi_k = n_k \frac{\pi}{N}, \quad n_k \in {\rm Z}.
\label{4.1}
\end{equation}
Nothing changes in the above solution of classical affine Toda field theory, except that the input parameters $\varphi_k$ are restricted to a set of discrete values.
The chiral twist of the $k$-th elementary soliton then becomes an element of the cyclic group Z$_N$, 
\begin{equation}
e^{-2i\varphi_k}=1^{n_k/N}\in {\rm Z}_N.
\label{4.2}
\end{equation}
As a consequence,
\begin{equation}
\left( \zeta_k \right)^N = (-1)^{N+n_k} e^{-N \xi_k},
\label{4.3}
\end{equation}
so that matrix elements of $\omega_n$ satisfy the relation
\begin{equation}
(\omega_{n+N})_{k\ell} = (-1)^{n_k+n_{\ell}} (\omega_n)_{k\ell} e^{N(\xi_{\ell}-\xi_k)} \quad ({\rm no\ sum}).
\label{4.4}
\end{equation}
Diagonal matrix elements in particular become periodic in the index $n$ with period $N$,
\begin{equation}
(\omega_{n+N})_{kk} = (\omega_n)_{kk}         \quad \quad ({\rm no\ sum}).
\label{4.5}
\end{equation}
In the present section, we focus on diagonal matrices $\omega_0$, corresponding to (multi-)soliton bound state and scattering solutions without breathers.
In this case, the set of matrices $\omega_n$ is periodic with period $N$,
\begin{equation}
\omega_{n+N} = \omega_n.
\label{4.6}
\end{equation} 
According to Eq.~(\ref{3.8}), this periodicity is inherited by $\psi_n$ and $\Phi_n$,
\begin{eqnarray}
\psi_{n+N} & = & \psi_n,
\nonumber \\
\Phi_{n+N} & = & \Phi_n.
\label{4.7}
\end{eqnarray}
As a result, the infinite set of Toda field equations (\ref{3.14}) can be truncated to a finite set of $N$ equations,
\begin{eqnarray}
- 4 \partial \bar{\partial} \Phi_n & = &  e^{\Phi_{n+1}-\Phi_n} - e^{\Phi_n- \Phi_{n-1}}  \quad (n=1,...,N),
\nonumber \\
\Phi_0 & = & \Phi_N, \quad \Phi_{N+1} = \Phi_1.
\label{4.8}
\end{eqnarray} 
These are the field equations for the affine Toda field theory corresponding to the Lie algebra su($N$) or A$_{N-1}^{(1)}$.
Likewise, the Lax representation can be reduced to an $N$-component vector $\psi$ with components $\psi_n, n=1,...,N$
and $n\times n$ matrices ${\cal A},{\cal B}$. Equations (\ref{3.11}-\ref{3.13}) remain valid with the restriction 
$n=1,...,N$, together with the cyclic conditions
\begin{equation}
\psi_0  =  \psi_N, \quad \psi_{N+1} = \psi_1, \quad \Phi_0 = \Phi_N, \quad \Phi_{N+1}= \Phi_1.
\label{4.9}
\end{equation}
As usual, one can impose one further condition
\begin{equation}
\sum_{n=1}^N \Phi_n = 0,
\label{4.10}
\end{equation}
since the sum of the field equations (\ref{4.8}) shows that the left hand side of (\ref{4.10}) is a free field.
Eq.~(\ref{4.8}) then gets replaced by the following set of $N-1$ equations,
\begin{eqnarray}
- 4 \partial \bar{\partial} \Phi_n & = &  e^{\Phi_{n+1}-\Phi_n} - e^{\Phi_n- \Phi_{n-1}}  \quad (n=1,...,N-1),
\nonumber \\
\Phi_0 & = & \Phi_N = - \sum_{n=1}^{N-1} \Phi_n.
\label{4.11}
\end{eqnarray} 
The Lax pair remains unchanged.

We have not yet exploited the ``reality condition" (\ref{3.15}). Supplemented by the periodicity condition (\ref{4.7}), it implies
\begin{equation}
\Phi_n = - \Phi_{N+1-n}^*
\label{4.12}
\end{equation}
and thus enables us to eliminate half of the bosonic fields. 

{$\bullet$ \bf even $N$:}

The independent fields are $\Phi_1,...,\Phi_{N/2}$, subject to the restriction that $\sum_{k=1}^{N/2}\Phi_k$ is real.
The total number of independent real fields is $N-1$. The field equations are
\begin{eqnarray}
-4 \partial \bar{\partial} \Phi_1 & = &  e^{\Phi_2-\Phi_1}-  e^{\Phi_1+\Phi_1^*} ,
\nonumber \\
-4 \partial \bar{\partial} \Phi_n & = &    e^{\Phi_{n+1}-\Phi_n} - e^{\Phi_n-\Phi_{n-1}}, \quad (n=2,..,N/2-1) \quad {\rm for\ } N \ge 6,
\nonumber \\
-4 \partial \bar{\partial} \Phi_{N/2} & = &  e^{-\Phi_{N/2} -\Phi_{N/2}^*}   - e^{\Phi_{N/2}-\Phi_{N/2-1}} \quad {\rm for\ } N \ge 4 .
\label{4.13}
\end{eqnarray}

{$\bullet$ \bf odd $N$:}

The independent fields are $\Phi_1, ..., \Phi_{(N-1)/2}$, again equivalent to $N-1$ real fields.
The field equations read
\begin{eqnarray}
 - 4 \partial \bar{\partial} \Phi_1 & = &   e^{\Phi_2-\Phi_1}  -e^{\Phi_1+\Phi_1^*} ,
\nonumber \\
 -  4 \partial \bar{\partial} \Phi_n & = &   e^{\Phi_{n+1}-\Phi_n}   -e^{\Phi_n-\Phi_{n-1}}, \quad (n=2,..,(N-1)/2-1) \quad {\rm for\ } N \ge 7,
\nonumber \\
 - 4 \partial \bar{\partial} \Phi_{(N-1)/2} & = &  e^{\sum_{k=1}^{(N-1)/2} (\Phi_k^*-\Phi_k) - \Phi_{(N-1)/2}} -  e^{\Phi_{(N-1)/2}-\Phi_{(N-1)/2-1}} \quad {\rm for\ } N \ge 5.
\label{4.14}
\end{eqnarray}
It may be worthwhile to consider the lowest few values of $N$ in greater detail:

$N=2$

$\Phi_1$ is real and satisfies the sinh-Gordon equation (\ref{1.6}) 
\begin{equation}
-4 \partial \bar{\partial}\Phi_1 =  e^{-2\Phi_1}  -e^{2\Phi_1} = -2 \sinh (2\Phi_1) , \quad (\Phi_1 = \theta/2).
\label{4.15}
\end{equation}
Only chiral twists of $(-1)$ are allowed, corresponding to the twist angles $\varphi_k=\pi/2$. We thus recover the known relationship
between kinks in the large $N$ GN model (or the classical $N=2$ GN model) and the sinh-Gordon model \cite{11,12}.
The Lax connection in the present  formalism is found to be
\begin{equation}
{\cal A}  =  \frac{1}{2i\zeta} \left( \begin{array}{cc} 0 & e^{-2\Phi_1} \\ e^{2\Phi_1} & 0 \end{array} \right),
\quad {\cal B} = \frac{1}{2i} \left( \begin{array}{cc} -2i \bar{\partial} \Phi_1 & - \zeta \\ - \zeta & 2i \bar{\partial} \Phi_1 \end{array} \right).
\label{4.16}
\end{equation}
It agrees with Eq.~(\ref{1.8}) modulo a gauge transformation. The relevant gauge transformation can be inferred from Eq.~(\ref{3.10}) and 
the periodicity conditions in $n$. It is easy to verify that
\begin{eqnarray}
{\cal A}_{\rm Eq. (9)} & = & \Omega^{-1} \left( {\cal A}_{\rm Eq. (61)} - \partial \right) \Omega,
\nonumber \\
{\cal B}_{\rm Eq. (9)} & = & \Omega^{-1} \left( {\cal B}_{\rm Eq. (61)}  - \bar{\partial} \right) \Omega,
\label{4.17}
\end{eqnarray}
with 
\begin{equation}
\Omega = \left( \begin{array}{cc} 0 & e^{-\Phi_1} \\ - \zeta^{-1} & 0 \end{array} \right), \quad \Phi_1 = \theta/2.
\label{4.18}
\end{equation}

$N=3$

There is one complex field $\Phi_1$,
\begin{equation}
-4 \partial \bar{\partial} \Phi_1 =   e^{\Phi_1^*-2 \Phi_1}  - e^{\Phi_1+\Phi_1^*}.
\label{4.19}
\end{equation}
For real $\Phi_1$, we recover the integrable Bullough-Dodd equation \cite{20}
\begin{equation}
- 4 \partial \bar{\partial} \Phi_1 =   e^{-\Phi_1} - e^{2\Phi_1} .
\label{4.20}
\end{equation}
For complex $\Phi_1 = \theta+i \tilde{\phi}$,  the field equations
\begin{eqnarray}
- 4 \partial \bar{\partial} \theta & = &  e^{-\theta} \cos 3 \tilde{\phi} - e^{2\theta} ,
\nonumber \\
- 4 \partial \bar{\partial} \tilde{\phi} & = & - e^{- \theta} \sin 3 \tilde{\phi} ,
\label{4.21}
\end{eqnarray}
agree with Eq.~(3.14) in a work by Fordy and Gibbons \cite{21}.

$N=4$

There are two complex fields $\Phi_1,\Phi_2$,
\begin{eqnarray}
- 4 \partial \bar{\partial} \Phi_1 & = &   e^{\Phi_2-\Phi_1}  - e^{\Phi_1+\Phi_1^*},
\nonumber \\
- 4 \partial \bar{\partial} \Phi_2 & = &   e^{-\Phi_2 - \Phi_2^*} - e^{\Phi_2-\Phi_1}. 
\label{4.23}
\end{eqnarray}
$\Phi_1+\Phi_2$ is real, leaving us with three independent real fields.

Results for larger $N$ can easily be obtained from the above general formalism.

\section{Soliton and breather solutions for finite $N$}
\label{sect5}

In the preceding sections, we assumed $\omega_0$ to be diagonal. This is sufficient to describe all purely solitonic solutions of the Dirac equation where
the building blocks --- elementary kinks or bound states of several kinks --- are static in their rest frame. In order to include also breathers with an
intrinsic time dependence in the rest frame, it is necessary to allow for off-diagonal matrices $\omega_0$. We recall the following observation from the study 
of transparent Dirac potentials \cite{2,9}: In order to avoid problems with cluster separability, solitons connected 
by off-diagonal matrix elements of $\omega_0$ must move with the same velocity. 
If one accepts this restriction of input parameters, Eq.~(\ref{4.4}) reduces to
\begin{equation}
(\omega_{n+N})_{k\ell} = (-1)^{n_k+n_{\ell}} (\omega_n)_{k\ell} \quad ({\rm no\ sum}) .
\label{5.1}
\end{equation}
If a particular solution containing breathers is such that the matrix $\omega_0$ only has off-diagonal elements $(\omega_0)_{k \ell}$ for which $n_k + n_{\ell}$ is even,
the matrices $\omega_n$ satisfy the same periodicity condition $\omega_{n+N}=\omega_n$ as for pure solitonic solutions. Then we can apply everything which 
has been said in Sect.~\ref{sect4}. In the most general case where $n_k+n_{\ell}$ is odd for some $(\omega_0)_{k\ell} \neq 0$, the period is simply doubled,
\begin{equation}
\omega_{n+2N} = \omega_n  .
\label{5.2}
\end{equation} 
In this case, one can use all the formulas above for even $N$ only. 

We illustrate this procedure for the simplest case, i.e., the su(3) affine Toda theory (there is no breather in the su(2) case in our formalism).
Since now $\Phi_n = \Phi_{n+6}$, we have to use Eqs.~(\ref{4.13}) for $N=6$,
\begin{eqnarray}
-4 \partial \bar{\partial} \Phi_1 & = & e^{\Phi_2-\Phi_1} -e^{\Phi_1 + \Phi_1^*},
\nonumber \\
-4 \partial \bar{\partial} \Phi_2 & = & e^{\Phi_3-\Phi_2} -e^{\Phi_2 - \Phi_1},
\nonumber \\
-4 \partial \bar{\partial} \Phi_3 & = & e^{-\Phi_3-\Phi_3^*} -e^{\Phi_3 - \Phi_2}.
\label{5.3}
\end{eqnarray} 
$\Phi_1,\Phi_2,\Phi_3$ are in general complex, but $\Phi_1+\Phi_2+\Phi_3$ is real. These equations are identical to those of the purely solitonic case for su(6).
The fact that we are dealing with su(3) rather than su(6) here will be encoded in the twist angles, quantized in steps of $\pi/3$, not $\pi/6$.

\section{Comparison with other works and outlook}
\label{sect6}

There is a considerable amount of literature on affine Toda field theories. Here, we only select some examples dealing
with classical solitons and breathers of the
su($N$) affine Toda model which can be compared directly to our results. Let us start with pure soliton solutions, pioneered by Hollowood \cite{22} (see also \cite{22a,22b}). The work about
multi-soliton solutions which matches our notation most closely is Ref.~\cite{23}. Expressions corresponding to our $e^{\Phi_n}$ in terms
of ratios of determinants can be found in Eqs. (11.43,11.44) of that paper. The matrix $V$ obviously corresponds to our matrix $B$, whereas $\Omega$ plays the role
of our $Z$. We note that the rapidity dependence of $Z$, Eqs.~(\ref{2.4},\ref{2.5}),  drops out when computing the determinants. We could have set all $\xi_k=0$ 
in the definition of $\omega_n$  from the beginning, so that $Z$ would become unitary, $(Z^{\dagger})^{-1}=Z$. We then recognize that our Eq.~(\ref{3.8})
and Eq.~(11.44) of Ref.~\cite{23} have identical structure. 

\begin{figure}[h]
\begin{center}
\epsfig{file=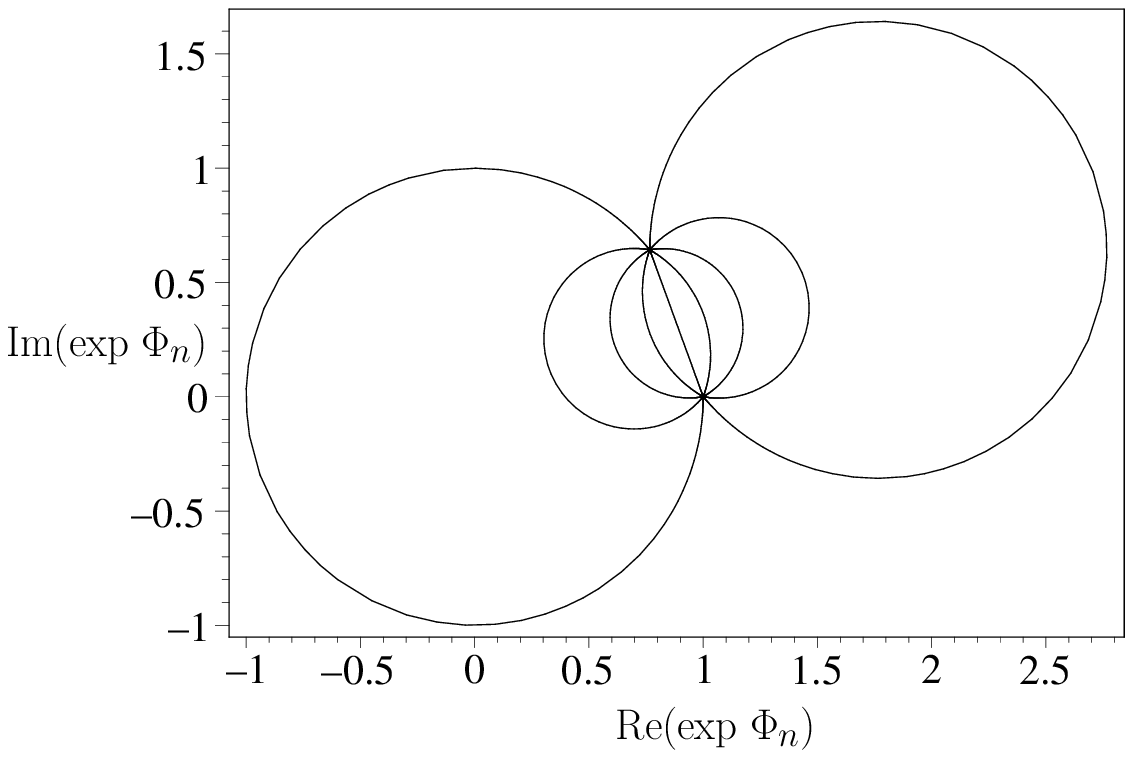,width=8cm,angle=0}
\caption{Illustration of potentials $e^{\Phi_n}, n=1,...,N$ for the single su($N$) Toda kink in the complex plane. The parameters chosen are $N=9, \varphi=\pi/N$. The straight
line segment represents the Dirac potential of the Shei kink, used as starting point for the whole construction.}
\label{fig1}
\end{center}
\end{figure}

By way of illustration, consider the single soliton solution at rest where 
\begin{equation}
e^{\Phi_n} = \frac{e^{2in\varphi} + U}{e^{2i(n-1)\varphi} + U}, \quad U = e^{2 x \sin \varphi},
\label{6.1}
\end{equation}
for the choice $\omega_0=(2 \sin \varphi)^{-1}$. For $n=1$, this is the Dirac potential for the single twisted kink which traces out a straight line
segment between the points $e^{2i\varphi}$ and 1 in the complex plane. For all other values of $n$, Eq.~(\ref{6.1}) represents a circular arc joining the 
same two points, as can be seen from the identity
\begin{equation}
e^{\Phi_n} = \frac{\sin((2n-1)\varphi)}{\sin(2(n-1)\varphi)} e^{i\varphi} - \frac{\sin \varphi}{\sin(2(n-1)\varphi)} \frac{e^{-2i(n-1)\varphi}+U}
{e^{2i(n-1)\varphi}+U} e^{i(2n-1)\varphi}.
\label{6.2}
\end{equation}
The expression on the right hand side is singular at $n=1$, since here one attempts to parametrize a straight line segment as an arc of a circle with infinite radius.
An example is shown in Fig.~\ref{fig1} for $N=9, \varphi=\pi/N$. The reflection symmetry with respect to the straight line segment is due to condition (\ref{3.15}).
For even $N$, one of the curves becomes ill-defined because the denominator vanishes at some point $x$. The way such singularities are avoided in Toda
models is by choosing a complex value of $\omega_0$, in our language. This breaks the symmetry (\ref{3.15}) and destroys the direct relation to the Dirac 
equation for $n=1$, but is of course perfectly legitimate within the Toda model. In Fig.~\ref{fig1}, this would manifest itself through the loss of the above mentioned reflection
symmetry and disappearance of the straight line. Nevertheless, all curves would remain circular. Note that already in the sinh Gordon model, 
Dirac kink potentials give rise to singular solitons of the bosonic theory, if the original parameters are kept. 

It is also instructive to apply Eq.~(\ref{6.1}) to the su(3) theory, i.e., Eqs.~(\ref{4.19},\ref{4.20},\ref{4.21}) above. We have to quantize the twist angle in steps
of $\pi/3$, yielding
\begin{eqnarray}
e^{\Phi_1} & = & \frac{e^{2i\pi/3} + e^{x \sqrt{3}}}{1+e^{x\sqrt{3}}} \quad (\phi=\pi/3),
\nonumber \\
e^{\Phi_1} & = & \frac{e^{-2i\pi/3} + e^{x\sqrt{3}}}{1+ e^{x\sqrt{3}}} \quad (\phi=2\pi/3).
\label{6.2a}
\end{eqnarray}
These two soliton solutions are complex conjugates. The value $\phi=3\pi/3$ is equivalent to $\phi=0$ and does not yield any soliton, but the vacuum $e^{\Phi_1}=1$.
In the notation of Fordy and Gibbons, Eq.~(\ref{4.21}), the two non-trivial solitons read
\begin{eqnarray}
\theta & = & \frac{1}{2} \ln \left( \frac{1-e^{x\sqrt{3}}+e^{2x\sqrt{3}}}{(1+e^{x\sqrt{3}})^2}\right),
\nonumber   \\
\tilde{\phi} & = & \pm i \arctan \left( \frac{\sqrt{3}}{2e^{x\sqrt{3}}-1} \right).
\label{6.2b}
\end{eqnarray}
We do not get any real soliton solution which would solve the Bullough-Dodd equation (\ref{4.19}).

Breather solutions have been explored in less detail in affine Toda field theory. The authors of Ref.~\cite{24} study various single breathers in the su($N$)
case. 
Let us compare the single breather at rest generated in our formalism to their result. To this end, we first note that the ``$\tau$-functions"
used in Ref.~\cite{24} correspond to 
\begin{equation}
\tau_n = \frac{{\rm det} (\omega_n + B)}{{\rm det} (\omega_n)}.
\label{6.2c}
\end{equation}
The denominator normalizes $\tau_n$ to 1 at $x \to -\infty$. We then set up the (2-soliton) breather at rest in our framework and try to match the expressions
for the $\tau$-functions given in \cite{24}. We find the following parameters,
\begin{eqnarray}
\theta_1 & = & - \theta_a/2 - \arctan v
\nonumber \\
\theta_2 & = &  -\theta_a/2 + \arctan v
\nonumber \\
\omega_{11} & = &  \omega_{22} = 0  
\nonumber \\
\omega_{12} & = & \omega_{21}^* = \frac{1+iv}{(1+v^2)\sigma_a} e^{-\rho}
\nonumber \\
\sigma_a & = & - \frac{2}{\sqrt{1+v^2}} \sin \frac{\theta_a}{2}
\label{6.2d}
\end{eqnarray} 
The resulting $\tau$-function,
\begin{eqnarray}
\tau_n & = &  1 + e^{\sigma_a(x+ivt)+\rho + in \theta_a} + e^{\sigma_a(x-ivt) +\rho +  i n \theta_a} + A e^{2 \sigma_a x + 2\rho + 2in \theta_a}
\nonumber \\
A & = &  \left( 1 - \frac{(1+v^2)^2 \sigma_a}{4 v^2} \right)^{-1}
\label{6.2e}
\end{eqnarray}
agrees with the ``type A breather" of \cite{24}, see the equation following Eq.~(4.17) of that work, provided we choose $\rho'=\rho$ and
certain relations among $\sigma_a, A, v$ and $\theta_a$. We should point out that the choice $\omega_{11}=\omega_{22}=0$ is not
allowed when constructing solutions of the fermionic models, since it would lead to singular mean fields. Apparently, we can obtain
the same functional form as in the Toda studies, but physical conditions on solutions in either the fermionic or the 
bosonic models are different. Hence one is not allowed to use the same parameters. We were not able to generate ``type B breathers" 
of Ref.~\cite{24} at all with our methods. On the other hand, we have more freedom in constructing more complicated breather candidates by allowing for diagonal
elements of $\omega$. This leads to $\tau$-functions involving 5 different exponentials rather than 3, which have not been considered in Ref.~\cite{24}.
Here, the main issue will be whether one can find parameter sets which avoid singularities in all components $\Phi_n$, but this is beyond the scope of
the present work. 

The most intriguing question raised by our study is the following: What is the physics behind the su($N$) symmetry? The original Dirac equation
with transparent scalar and pseudoscalar potential does not give much of a clue. If we consider the relationship between these potentials and 
self-consistent fields of fermionic models like the GN and NJL$_2$ models, there is an obvious candidate:
These models were considered in the large $N$ limit of SU($N$) flavor symmetry. If we recall the self-consistency condition for the twisted kink,
the argument becomes even more compelling. The relationship between the twist angle $\varphi_k$  and the occupation of the valence bound state
is 
\begin{equation}
n_k = \frac{\varphi_k}{\pi} N
\label{6.3}
\end{equation}
We may take mean field theory as an approximation to finite $N$ GN and NJL$_2$ models rather than the large $N$ limit, similar to what is routinely done in condensed
matter physics. Then, $n_k$ can only assume the values 1,...,$N$ due to the Pauli principle, so that $\varphi_k$ gets discretized in steps of $\pi/N$. But this was
exactly what it took to proceed from su($\infty$) to su($N$) Toda models for finite $N$. 
This strongly suggests a relationship between SU($N$) flavor symmetry of fermionic models and the su($N$) Lie algebra underlying bosonic Toda models
which deserves further study. 

Finally, we remind the reader that the works on transparent Dirac potentials have recently been generalized to matrix potentials $\Delta$, i.e., coupled Dirac
equations. This was done in two different physics settings, first in the context of the multi-component BdG equation for exotic superconductors \cite{25,26}
and later for solving the NJL$_2$ model with flavor and color \cite{27,28}. Since this requires only a mild generalization of the formalism, it may be interesting
to repeat the present investigation for the multi-component case and hopefully end up with an integrable matrix generalization of affine Toda field theory.

 

\end{document}